\documentclass[preprint,amsmath,amssymb,aps,floatfix,pre,superscriptaddress]{revtex4-1}
\usepackage{graphicx}
\usepackage{pifont}
\begin{document}

\title{Collective Modes in Two Dimensional Binary Yukawa Systems}

\author{Gabor J. Kalman}
\email{kalman@bc.edu}
\affiliation{Department of Physics, Boston College, Chestnut Hill, MA 02467, USA}

\author{Peter Hartmann}
\affiliation{Institute for Solid State Physics and Optics, Wigner Research Centre for Physics, Hungarian Academy of Sciences, P.O.B. 49, H-1525 Budapest, Hungary}
\affiliation{Department of Physics, Boston College, Chestnut Hill, MA 02467, USA}

\author{Zolt\'an Donk\'o}
\affiliation{Institute for Solid State Physics and Optics, Wigner Research Centre for Physics, Hungarian Academy of Sciences, P.O.B. 49, H-1525 Budapest, Hungary}
\affiliation{Department of Physics, Boston College, Chestnut Hill, MA 02467, USA}

\author{Kenneth I. Golden}
\affiliation{Department of Mathematics and Statistics and Department of Physics, University of Vermont, Burlington, VT 05401-1455, USA}

\author{Stamatios Kyrkos}
\affiliation{Department of Chemistry \& Physics, Le Moyne College, 1419 Salt Springs Road, Syracuse, NY 13214, USA}

\date{\today}
\begin{abstract}

We analyze via theoretical approaches and molecular dynamics simulations the collective mode structure of strongly coupled two-dimensional binary Yukawa systems, for selected density, mass and charge ratios, both in the liquid and crystalline solid phases. Theoretically, the liquid phase is described through the Quasi-Localized Charge Approximation (QLCA) approach, while in the crystalline phase we study the centered honeycomb and the staggered rectangular crystal structures through the standard harmonic phonon approximation. We identify ``longitudinal'' and ``transverse'' acoustic and optic modes and find that the longitudinal acoustic mode evolves from its weakly coupled counterpart in a discontinuous non-perturbative fashion. The low frequency acoustic excitations are governed by the oscillation frequency of the average atom, while the high frequency optic excitation frequencies are related to the Einstein frequencies of the systems.

\end{abstract}


\maketitle

\section{Introduction}

Yukawa systems, i.e. many particle systems where the pair interaction potential energy is 
\begin{eqnarray}
\phi(r)&=&Z_1Z_2\varphi(r) \nonumber \\
\varphi(r)&=&\frac{\exp(-\kappa r)}{r}
\end{eqnarray} have been  of interest for some  time. The Yukawa potential has the unique feature that by varying the screening parameter $\kappa$ the potential can assume the feature both of a short range (hard sphere like) and of a long range (Coulomb like) interaction potential. Since the 1970-s this feature motivated a number of investigations relating to the properties of Yukawa liquids and solids, their phases and phase transitions \cite{LOWEN12,LOWEN15,LOWEN16}. Quite apart from this academic interest, the Yukawa potential has been recognized as a good approximation for the interaction potential between charged particles in colloids \cite{LOWEN} and, more recently, in complex (dusty) plasmas, where the original Coulomb interaction between the main constituents is transformed by Debye screening into a Yukawa-type potential (for a review see, e.g. \cite{MorfREV,BonREV}). Complex plasmas constitute an especially suitable medium for the study of waves and collective excitations because these are  much less damped than in colloidal systems. In recent years a host of papers, both theoretical \cite{WangPRL01,PeetersPRA87,MurilloPRL00,DonkoJPC,PRL-Einstein,HartmannJPA06,IEEED,GoldenPRE10,KalmanEPL10} and experimental \cite{NosenkoPRL06,PielPOP06} have studied collective modes in Yukawa systems \cite{SmithASR08,MatthewsASR06,QiaoPRE03,RaoPSS90,RaoPSS90,MelzerPRE03,LinI00,LinI01,Fortov04}.
Most of the experimental work on colloidal systems and complex plasmas has focused attention on two dimensional (2D) layers. We note that the physics of the 2D and 3D systems, the dynamics of the collective excitations in particular, is, in fact, quite different and addressing 2D and 3D systems separately is also warranted on theoretical grounds \cite{LOWEN}.

The strength of the coupling between the particles can be characterized by the nominal bare coupling parameter $\Gamma=(Z^2e^2)/(k_{\rm b}Ta)$, with $a$ being the Wigner-Seitz radius. A physically more meaningful $\Gamma_{\rm eff}$ that, basically represents the ratio of the potential and kinetic energies, can be defined for orientation purposes as $\Gamma_{\rm eff}=\Gamma\exp(-\kappa a)$ \cite{IkeziPF86}, although more sophisticated expressions are available \cite{Vau02,HaPRE05,OtteffG}. 

The main interest lies in the behavior of the strongly coupled state, $\Gamma_{\rm eff} \gg 1$. In this strongly coupled state the system can be either in the dense liquid or in the crystalline solid phase.

Past works on the dynamics have overwhelmingly concentrated on Yukawa systems consisting of one single component, the equivalent of the One Component Plasma, both in three and two dimensions (Y3dOCP and Y2dOCP, respectively). A great deal of theoretical \cite{MurilloPRL00,MurilloPOP98,MurilloPOP00,PRL-Einstein,HartmannJPA06,IEEED,GoldenPRE10,KalmanEPL10} and computer simulation \cite{HamaguchiPRE97,OhtaPRL00,DonkoJPC,PPL10,DonkoJPA09} efforts have been devoted to the mapping and understanding of the collective mode structures in these systems, both in the liquid and solid phases. The theoretical methods required in the two situations are, of course, quite different. Once the lattice structure is identified, the crystalline solid is amenable to the standard harmonic phonon analysis. Concerning the treatment of the collective modes in the strongly coupled liquid phase, the Quasi Localized Charge Approximation (QLCA) approach developed by Kalman and Golden \cite{KalmanPRA90,GoldenPOP00} has turned out quite successful. The QLCA borrows ideas from the harmonic phonon approximation and describes the liquid in terms of particles trapped  and oscillating in local potential wells of the fluctuating potential (for details, see \cite{GoldenPOP00}). As a result of these works, the collective mode spectra of the Y3dOCP and Y2dOCP are well understood and this understanding is well corroborated by observations \cite{NosenkoPRL06,PielPOP06}.  

As to strongly coupled Yukawa mixtures consisting of more than one single species, in particular binary Yukawa mixtures (Y3dBM, Y2dBM), the collective dynamics of these systems constitutes a largely unexplored area (see, however, a recent work by Daligault \cite{DaligaultPRL12}) , even though the problem is of great theoretical interest. (For a related one dimensional problem see \cite{FerreiraPRB08,FerreiraJPC10}). One expects that the simple analytic structure of the Yukawa potential allows one to derive nearly exact solutions, which will elucidate the common features of the dynamics of binary liquids and solids \cite{ScaliseJCE09,ScaliseFPE10}. Also, the flexibility of the Yukawa interaction would make the qualitative features of the results to serve as paradigms for the collective mode structures of binary systems interacting through other potentials as well (alloys, dipole systems, etc.). From the point of view of actual applications, the creations of binary complex plasmas has technical problems, but, nevertheless, one expects that such strongly coupled complex plasmas of two different grain species will become available in the near future.  

This paper, the first in a series, presents a systematic study of he collective mode spectra of the Y2dBM system. The system consists of two species, with charges $Z_1$ and $Z_2$, masses $m_1$ and $m_2$ and densities $n_1$ and $n_2$ (or concentrations $c_1$ and $c_2$), respectively. Our strategy is similar to the one followed in our previous works on the Y3dOCP and Y2dOCP: for the theoretical analysis of the liquid state we apply the QLCA formalism; for the crystalline solid we calculate dispersion relations by the standard method. In both cases, we parallel our theoretical analysis with detailed Molecular Dynamics (MD) simulations of the density and current fluctuation spectra of the system; it is, then, the positions of the peaks of the fluctuation spectra from which the dispersion relations are inferred. It has to be noted, however, that following this road map is fraught with questions stemming from the fundamental difference between the binary and single component systems. Concerning the QLCA, is it justified to represent the system through separate collective coordinates for each of the species in a liquid where the two species are spatially not separated? Concerning the MD, if different partial fluctuation spectra provide conflicting information, which one of them should be accepted as most relevant to the actual dispersion? Finally, one has to be aware of the fact that in the presence of different $Z_1$ and $Z_2$ charges with different $c_1$ and $c_2$ concentrations, the liquid phase is governed by a complex phase diagram \cite{ScaliseJCE09,ScaliseFPE10,OgataPRE93,IyetomiPRB89,LowenJPC12,JiangEPL11,JiangIEE12} in which only certain combinations of these parameters allows a homogeneous system. In the solid phase, similarly, with a given set of parameters only certain lattice structures are permissible \cite{LOWEN1,LOWEN2}.

We tackle these issues along the work as presented in the sequel. We have to emphasize though, that our goal is restricted to determining the existence, interrelationships and dispersion of the collective modes. We do not address a number of related problems: the damping of the modes, the detailed structures and the link between the various fluctuation spectra, the critical freezing values of $\Gamma$, the nature of the underlying order in the liquid phase, lattice stability and structures, etc. The issues investigated in this paper are organized according to the following plan: Section II is devoted to the description of the liquid phase and Section III of the crystalline solid phase. In each case we first analyze the qualitative features of the optic ($\omega$ finite at $k=0$) and then the acoustic ($\omega \rightarrow 0$ as $k \rightarrow 0$) excitations, before presenting the description of the full mode structure. In Section III we compare the mode structures in the two phases and draw conclusions. (For a preliminary account of some of the results pertaining to the optic modes see \cite{KalmanCPP12} and to the acoustic modes see \cite{PRL-2011}). 

Whenever not noted otherwise, we measure frequencies in units of $\omega_1$, the plasma frequency of species 1, use $\Gamma_1$, the bare coupling value for species 1 to characterize the coupling strength, and adopt $\kappa a=1$ ($a=\sqrt{a_1 a_2}$) for the screening parameter.

\section{Strongly Coupled Liquid Phase}

The theoretical analysis of the mode structure in the liquid state is based on the QLCA approach, as discussed above. The fundamental equation for the dynamical matrix is
\begin{eqnarray}
C_{AB}^{\mu\nu}({\bf k})&=&-Z_AZ_Be^2\frac{\sqrt{n_An_B}}{\sqrt{m_Am_B}}\Bigg[\int {\rm d}^2r \Bigg\{\Psi^{\mu\nu}({\bf r})\left(\exp(-i{\bf k}\cdot {\bf r})-\delta_{AB}\right)\left[1+h_{AB}(r)\right]- \nonumber \\ 
& &\delta_{AB}\sum_{C\ne A}\frac{Z_Cn_C}{Z_An_A}\Psi^{\mu\nu}({\bf r})\left[1+h_{AC}(r)\right] \Bigg\} \Bigg] \nonumber \\
&=& -\omega^2_{AB}\frac{1}{2\pi}\int{\rm d}^2\bar{r}\Psi^{\mu\nu}({\bf \bar{r}})\left(\exp(-i{\bf k}\cdot {\bf r})-\delta_{AB}\right)\left[1+h_{AB}(\bar{r})\right]+ \nonumber \\
& & \delta_{AB}\sum_{C\ne A}\Omega^2_{AC}\int{\rm d}^2\bar{r}\Psi^{\mu\nu}({\bf \bar{r}})\left[1+h_{AC}(\bar{r})\right]
\end{eqnarray}
with
\begin{equation}
\Psi^{\mu\nu}(r) = \partial_\mu \partial_\nu \varphi(r),
\end{equation}
where $\varphi(r)$ is the Yukawa interaction, $\varphi(r)=\exp(-\kappa r)/r$ characterized by the screening constant $\kappa$. Then
\begin{eqnarray}
\Psi^{\mu\nu}(r) &=& \frac{\exp(-\kappa r)}{r}\left(3\frac{r^\mu r^\nu}{r^2} a(\kappa r)-\delta^{\mu \nu} b (\kappa r)\right) \nonumber \\
a(y) &=& 1+y+\frac13 y^2, ~~~~~~~~b(y)=1+y.
\end{eqnarray}

Additional notational conventions are
\begin{eqnarray}
\Omega^2_{AB} &=&\frac{2\pi e^2 Z_AZ_Bn_B}{m_A a} \nonumber \\
\omega^2_{AB} &=&\frac{2\pi e^2 Z_AZ_B\sqrt{n_An_B}}{\sqrt{m_Am_B} a} \nonumber \\
a &=& \sqrt{a_1a_2} \nonumber \\
a_A &=& 1/\sqrt{\pi n_A} \nonumber \\
\bar{\kappa} &=& \kappa a \nonumber \\
\bar{r} &=& r/a \nonumber \\
y &=& \kappa r
\end{eqnarray}
with $Z$, $m$, $n$ and $a$ representing the charge number, mass, density and Wigner-Seitz radius for the  respective components. The $\Omega_{AB}$ and $\omega_{AB}$ frequencies are the nominal Einstein and nominal plasma frequencies of the system. The $h_{AB}$ pair correlation functions are to be obtained from the MD simulations, as described below.
  
The elements of the $C$-matrix can be expressed in terms of the kernel functions  $\cal{K}$, $\cal{L}$:
\begin{eqnarray} \label{eq:kernels}
C^L_{AB} &=& \omega^2_{AB} \int\frac{{\rm d}\bar{r}}{\bar{r}^2}{\cal K}(kr,y)\left[1+h_{AB}(\bar{r})\right] - \nonumber \\ 
& & \delta_{AB}\sum_{C(all)}\Omega^2_{BC}\int\frac{{\rm d}\bar{r}}{\bar{r}^2}{\cal K}(0,y)\left[1+h_{BC}(\bar{r})\right] \nonumber \\
C^T_{AB} &=& \omega^2_{AB} \int\frac{{\rm d}\bar{r}}{\bar{r}^2}{\cal L}(kr,y)\left[1+h_{AB}(\bar{r})\right] - \nonumber \\
& & \delta_{AB}\sum_{C(all)}\Omega^2_{BC}\int\frac{{\rm d}\bar{r}}{\bar{r}^2}{\cal L}(0,y)\left[1+h_{BC}(\bar{r})\right] 
\end{eqnarray}
with the kernel functions given by
\begin{eqnarray}
{\cal K}(u,r) &=& -\exp(-y)\left\{\left[1+y+y^2\right]J_0(u)-3\left[1+y+y^2/3\right]J_2(u) \right\} \nonumber \\
{\cal L}(u,r) &=& -\exp(-y)\left\{\left[1+y+y^2\right]J_0(u)+3\left[1+y+y^2/3\right]J_2(u) \right\} 
\end{eqnarray}

In order to clearly display the behavior in the vicinity of $k=0$
we also introduce
\begin{eqnarray}
{\cal G}(u,r) &=& {\cal K}(u,r)-{\cal K}(0,r) \nonumber \\
{\cal H}(u,r) &=& {\cal L}(u,r)-{\cal L}(0,r) \nonumber \\
{\cal F}(r) &=& -{\cal K}(0,r) = -{\cal L}(0,r)
\end{eqnarray}

The integrals of the kernel functions over the pair correlation functions $1+h(r)$ are
\begin{eqnarray} \label{eq:kerint}
K_{AB}(k) &=& \int\frac{{\rm d}\bar{r}}{\bar{r}^2}{\cal K}(kr,r)\left[1+h_{AB}(r)\right] \nonumber \\
L_{AB}(k) &=& \int\frac{{\rm d}\bar{r}}{\bar{r}^2}{\cal L}(kr,r)\left[1+h_{AB}(r)\right] \nonumber \\
F_{AB} &=& \int\frac{{\rm d}\bar{r}}{\bar{r}^2}{\cal F}(r)\left[1+h_{AB}(r)\right]
\end{eqnarray}

These integrals would be divergent at $r=0$, were it not for the pair correlation function $1+h(r)$ that becomes 0 at $r=0$. Similarly

\begin{eqnarray}
G_{AB}(k) &=& \int\frac{{\rm d}\bar{r}}{\bar{r}^2}{\cal G}(kr,r)\left[1+h_{AB}(r)\right] \nonumber \\
H_{AB}(k) &=& \int\frac{{\rm d}\bar{r}}{\bar{r}^2}{\cal H}(kr,r)\left[1+h_{AB}(r)\right]
\end{eqnarray}

Introducing the asymmetry parameters $p$ and $q$
\begin{eqnarray}
p^2 &=& Z_2n_2 / Z_1n_1 \nonumber \\
q^2 &=& Z_2m_1 / Z_1m_2
\end{eqnarray}
one obtains for the longitudinal elements
\begin{eqnarray} \label{eq:Clong}
C^L_{11}(k) &=& \frac{\omega_1^2}{2}\left[ G_{11}(k)+p^2F_{12} \right] \nonumber \\
C^L_{12}(k) &=& \frac{\omega_1^2}{2}pq\left[ G_{12}(k)-F_{12} \right] \nonumber \\
C^L_{22}(k) &=& \frac{\omega_1^2}{2}\left[ p^2q^2G_{22}(k)+q^2F_{12} \right] 
\end{eqnarray}
while the transverse elements are
\begin{eqnarray} \label{eq:Ctran}
C^T_{11}(k) &=& \frac{\omega_1^2}{2}\left[ H_{11}(k)+p^2F_{12} \right] \nonumber \\
C^T_{12}(k) &=& \frac{\omega_1^2}{2}pq\left[ H_{12}(k)-F_{12} \right] \nonumber \\
C^T_{22}(k) &=& \frac{\omega_1^2}{2}\left[ p^2q^2H_{22}(k)+q^2F_{12} \right]
\end{eqnarray}

We have found it useful to introduce $\omega_1$ ($=\omega_{11}$) as the reference frequency. In  general, there exist 4 modes as the roots of the characteristic equations, 
\begin{equation}
||C^{L,T}_{AB}-\omega^2|| = 0,
\end{equation}
which will be labeled $\omega^L_{+}$, $\omega^T_{+}$, $\omega^L_{-}$, $\omega^T_{-}$. The $\pm$ notation identifies the polarizations in species space of the modes: the ``+'' sign designates polarization where the two components move in-phase, while the ``--'' sign designates polarization where the two components move out-of-phase. The two + modes are acoustic ($\omega \rightarrow 0$ as $k \rightarrow 0$) and the two -- modes  are optic modes ($\omega$ finite for $k=0$). In addition, the modes are labeled as Longitudinal $L$ or Transverse $T$, referring to the their polarization with respect to $k$ when the propagation is along the principal axes. We note that the elements of the $C$-matrix, and consequently the eigenfrequencies, depend only on the two $p$ and $q$ combinations of the originally introduced three $Z=Z_2/Z_1$, $M=m_2/m_1$, $N=n_2/n_1$ parameters.

\subsection{Optic modes}

At $k=0$, $G_{AB}(k) \propto H_{AB}(k) \propto O(k^2)$, thus $\omega_{-}(k=0)$, the gap frequency, is longitudinal/transverse degenerate, as it should be for an isotropic liquid:
\begin{eqnarray} \label{eq:Qopt}
\omega_{\rm GAP} &=& \omega^L_{-}(k=0) = \omega^T_{-}(k=0) = \omega_1\sqrt{\frac12\left(p^2+q^2\right)F_{12}} \nonumber \\
&=& \sqrt{\frac12\left(\Omega_{12}^2+\Omega_{21}^2\right) F_{12}} = \sqrt{\frac12\left(\bar{\Omega}_{12}^2+\bar{\Omega}_{21}^2\right)}.
\end{eqnarray}
In view of Eqs. (\ref{eq:kernels}) through (\ref{eq:kerint}) $F_{AB}$ can be interpreted as the average potential generated by species $B$ in the environment of a particle of species $A$.
\begin{eqnarray} \label{eq:FFF}
F_{AB} &=& \frac{1}{2\pi}\int {\rm d}^2\bar{r}\langle\Psi(r)\rangle \left[1+h_{AB}(r)\right] \nonumber \\
\bar{\Omega}^2_{AB} &=& \Omega^2_{AB} F_{AB}
\end{eqnarray}
with $\langle \dots \rangle$ designating angular averaging. The $\bar{\Omega}_{AB}$ frequency represents the oscillation frequency of a particle of species $A$ in the frozen environment of particles of species $B$. We note that it is the correlation dependent $\bar{\Omega}$-s, rather than the nominal $\Omega$-s that are the real Einstein frequencies of the system \cite{PPL10}, with a similar definition being used in the theory of liquids \cite{CPP43}. In a single component system the Einstein frequency $\bar{\Omega}$ also provides the $\omega(k\rightarrow \infty)$ limiting frequency \cite{PRL-Einstein}. 
 
In order to find the $k\rightarrow \infty$ limits for the binary systems we re-express the elements of the $C$-matrix as
\begin{eqnarray}
C^L_{11} &=& \frac{\omega^2_{11}}{2}\left\{K_{11}(k)+\left(\bar{\Omega}^2_{11} + \bar{\Omega}^2_{12}\right)\right\} \nonumber \\
C^L_{12} &=& \frac{\omega^2_{12}}{2}K_{12}(k) \nonumber \\
C^L_{22} &=& \frac{\omega^2_{22}}{2}\left\{K_{22}(k)+\left(\bar{\Omega}^2_{22} + \bar{\Omega}^2_{21}\right)\right\}
\end{eqnarray}

In the $k\rightarrow \infty$ limits the $K$-terms vanish. This can be seen by observing that
\begin{equation}
K_{AB}(k) = k\int\frac{{\rm d}u}{u^2}{\cal K}(u,y)\left[1+h_{AB}(u/k)\right]
\end{equation}
and that $\left[1+h(r\rightarrow 0)\right] \rightarrow0$ fast enough to make this happen. Similar considerations apply to the transverse elements.

Thus the $k\rightarrow \infty$ respective upper and lower effective Einstein frequencies become $\Omega_{I}, \Omega_{II}$:
 
\begin{eqnarray} \label{eq:EinF}
\omega_{-}(k\rightarrow \infty) &=& \sqrt{\frac12 \left(\bar{\Omega}^2_{11} + \bar{\Omega}^2_{12}\right)} = \bar{\Omega}_{I} \nonumber \\
\omega_{+}(k\rightarrow \infty) &=& \sqrt{\frac12 \left(\bar{\Omega}^2_{22} + \bar{\Omega}^2_{21}\right)} = \bar{\Omega}_{II}. 
\end{eqnarray}

The calculated gap frequencies and the effective Einstein frequencies as functions of $\Gamma$, together with the results obtained by MD simulations (see below) are shown in FIG \ref{fig:GapEins}; also shown is the variation of the correlation integral $F_{12}$. In anticipation of the results of the next Section, we have also indicated the gap frequencies in the crystal lattices. We will further comment on the relationships between these gap frequencies in the next Section.

\begin{figure}[htbp]
\begin{center}
\includegraphics[width=8cm]{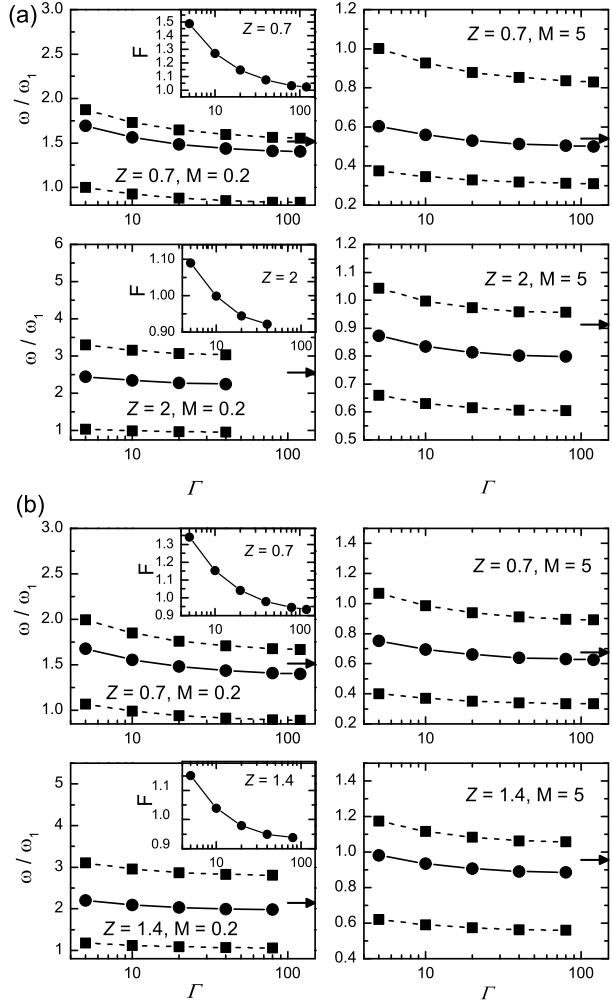}
\caption{Liquid state: QLCA gap (\ding{108}) and Einstein (\ding{110}) frequencies vs. $\Gamma$. The arrows indicate the positions of the corresponding gap in the lattice. The Inset shows the $\Gamma$ dependence of the correlation integral $F_{12}$ (symbols), which does not vary with the mass ratio. (a) $n_2 = n_1/2$; (b) $n_2=n_1$.}
\label{fig:GapEins}
\end{center}
\end{figure}

\subsection{Acoustic modes and sound speed}

We now turn to the calculation of the acoustic modes in the binary system. We are interested primarily in the small-$k$ behavior, which will lead to the determination of the sound speed.

First we observe that by dropping $h(r)$ in the integrals for the $G(k)$ and $H(k)$ functions the resulting $G^0(k)$ and $H^0(k)$ integrals become doable and provide the RPA expressions:
\begin{eqnarray} \label{eq:G0H0}
G^0(k) &=& \int\frac{{\rm d}\bar{r}}{\bar{r}^2}\exp(-\kappa r)\left[\left\{1+y+y^2\right\}\left\{1-J_0(u)\right\} - 3\left\{1+y+y^2/3\right\}J_2(u)\right] = \frac{\bar{k}}{\sqrt{\bar{\kappa}^2+\bar{k}^2}} \nonumber \\
H^0(k) &=& \int\frac{{\rm d}\bar{r}}{\bar{r}^2}\exp(-\kappa r)\left[\left\{1+y+y^2\right\}\left\{1-J_0(u)\right\} + 3\left\{1+y+y^2/3\right\}J_2(u)\right] = 0
\end{eqnarray}
The $C$-matrix equivalent to the cold RPA approximation would be obtained by dropping the $F_{12}$ terms in (\ref{eq:Clong}) and (\ref{eq:Ctran}), and using (\ref{eq:G0H0}) for $C_{11}$, $C_{12}$ and $C_{22}$. Then one obtains the RPA result
\begin{eqnarray}  
\omega_{+}^L &=& \omega_0\frac{\bar{k}}{\sqrt{\bar{\kappa}^2+\bar{k}^2}} \nonumber \\
\omega_{-}^L &=& 0 \nonumber \\
\omega_0 &=& \omega_1\sqrt{1+p^2q^2} = \sqrt{\omega_1^2+\omega_2^2}.
\label{eq:RPAw}
\end{eqnarray}

Note that the intuitively more reasonable requirement that in order to obtain the RPA limit one sets $h_{12}$ equal to zero everywhere in (\ref{eq:Clong}) and (\ref{eq:Ctran}) would result in a meaningless divergent integral for $F_{12}$. This feature shows that there is no smooth transition from the QLCA expression to the RPA. In other words, in contrast to the case of the YOCP, in the YBM the RPA Eq. (\ref{eq:G0H0}) cannot be simply amended by adding correlational corrections in order to obtain the strong coupling expression: the strong correlations show up in an essentially non-perturbative fashion.

Returning now to Eqs. (\ref{eq:Clong}) and (\ref{eq:Ctran}) we now calculate the small $k$ expansion. The result is given in terms of the integrals
\begin{eqnarray}
U_{AB} &=& -\frac{5}{16}\int_0^\infty {\rm d}y\left[1+y+\frac35y^2\right]\exp(-y)h_{AB}(r), \nonumber \\
V_{AB} &=& -\frac{1}{16}\int_0^\infty {\rm d}y\left[1+y-y^2\right]\exp(-y)h_{AB}(r).
\end{eqnarray}
Thus the longitudinal and transverse $C^L_{AB}$ and $C^T_{AB}$ matrix elements in the $k\rightarrow 0$ limit become
\begin{eqnarray} \label{eq:221}
C^L_{11}(k\rightarrow 0) &=& \frac{\omega_1^2}{2}\left\{(1-U_{11})\frac{\bar{k}^2}{\bar{\kappa}} + \frac12 p^2 \bar{\kappa} F_{12}\right\} \nonumber \\
C^L_{12}(k\rightarrow 0) &=& \frac{\omega_1^2}{2}\left\{pq(1-U_{12})\frac{\bar{k}^2}{\bar{\kappa}} - \frac12 pq \bar{\kappa} F_{12}\right\} \nonumber \\
C^L_{22}(k\rightarrow 0) &=& \frac{\omega_1^2}{2}\left\{p^2q^2(1-U_{22})\frac{\bar{k}^2}{\bar{\kappa}} + \frac12 q^2 \bar{\kappa} F_{12}\right\} \nonumber \\
C^T_{11}(k\rightarrow 0) &=& \frac{\omega_1^2}{2}\left\{V_{11}\frac{\bar{k}^2}{\bar{\kappa}} + p^2 \bar{\kappa} F_{12}\right\} \nonumber \\
C^T_{12}(k\rightarrow 0) &=& \frac{\omega_1^2}{2}\left\{pqV_{12}\frac{\bar{k}^2}{\bar{\kappa}} - \frac12 pq \bar{\kappa} F_{12}\right\} \nonumber \\
C^T_{22}(k\rightarrow 0) &=& \frac{\omega_1^2}{2}\left\{p^2q^2V_{22}\frac{\bar{k}^2}{\bar{\kappa}} + \frac12 q^2 \bar{\kappa} F_{12}\right\}.
\end{eqnarray} 

Proceeding now from (\ref{eq:221}), after some algebra one finds the small-$k$ expansion of the relevant $\omega_{+}^L(k)$, $\omega_{+}^T(k)$ mode frequencies as                                                                                                                                                                                                                                                                                                                                            
\begin{eqnarray}\label{eq:222}
(\omega_{+}^{L})^2(k\rightarrow 0) &=& \bar{\omega}^2\left\{1-\frac{U_{11}+2p^2U_{12}+p^4U_{22}}{\left(1+p^2\right)^2}\right\}\frac{\bar{k}^2}{\bar{\kappa}} \nonumber \\
(\omega_{+}^{T})^2(k\rightarrow 0) &=& \bar{\omega}^2\left\{\frac{V_{11}+2p^2V_{12}+p^4V_{22}}{\left(1+p^2\right)^2}\right\} \frac{\bar{k}^2}{\bar{\kappa}} 
\end{eqnarray}

While the first term in $(\omega_{+}^{L})^2 $ is RPA-like in appearance since it shows no explicit dependence on $h(r)$, in fact it reflects an essentially strong coupling behavior, the correlational effects manifesting themselves through the $\bar{\omega}$ coefficient, which we will refer to as the ``virtual average atom'' (VAA) frequency (this frequency has also been mentioned in relation to the self-diffusion coefficient of a plasma in \cite{HansenJoly}).
\begin{equation} \label{eq:bon}
\bar{\omega}^2=\omega_1^2\frac{q^2}{p^2+q^2}\left(1+p^2\right)^2.
\end{equation}
  
The Virtual Atom in fact represents an entity created from the averages of the system parameters. To see this, Eq. \ref{eq:bon} is re-written in terms of the average charge and mass as
\begin{eqnarray} \label{eq:226}
\bar{\omega} &=& \sqrt{\frac{2\pi e^2}{a}\frac{\langle Z\rangle^2}{\langle m \rangle}n}, \nonumber \\
n &=& n_1 + n_2.
\end{eqnarray}

The averages are defined through
\begin{equation}
\langle X \rangle = \frac{\sum_i X_in_i}{\sum_i n_i}.
\end{equation}

Compare now $\bar{\omega}$ with of Eq. \ref{eq:RPAw}: the dramatic difference in the dependence on the plasma parameters, in particular  on the mass ratio, is evident. (A similar result but restricted to the $Z_1=Z_2$ case was already anticipated in \cite{PRL-2011}).

The notion of the VAA originates from the literature, pertaining to liquid alloys and disordered binary systems \cite{VAA1,VAA2,VAA3}, as a heuristic concept. Here the derivation of this behavior, as a result of the evolution of the system from weak to strong coupling, is given.

All the observations now made on the $k\rightarrow 0$ behavior of the acoustic mode can be translated into 
statements about the sound speeds 
\begin{equation} \label{eq:225}
s^{L,T} = \left[\omega_{+}^{L,T}(k)/k\right]_{k\rightarrow 0}.
\end{equation}

Thus, according to (\ref{eq:221}) and (\ref{eq:225}), the longitudinal sound speed at weak coupling has its RPA value, governed by $\omega_0$; for strong correlations  the sound speed is substantially reduced and strong correlations manifest themselves, in contrast to the YOCP, in two ways: first, by morphing the mean field contribution  into one whose properties are dictated by the VAA and do not explicitly  depend on the correlations and, second, by generating an explicit correlational correction. For the transverse sound speed, similarly to the YOCP,  there is no $h$-independent contribution.

In parentheses we remark that to what extent the weak coupling value of the sound velocity is well represented by the RPA (or ``cold fluid'') expression is not clear. It is generally assumed that it is \cite{HansenJoly,MARLENE}. Nevertheless, the issue is that while for a Coulomb system there exists a clear rigorous derivation (also supported by ample observational evidence) that shows that in the $\Gamma \rightarrow 0$ limit the RPA is  correct, no such demonstration is currently available for a Yukawa system. In fact, there is reason to believe \cite{GoldenX} that for a finite range system the description of the behavior of the system in the weak coupling limit is more involved. All this, however, has very little bearing on our conclusion that the sound speeds and the low frequency excitations in the strongly coupled system are governed by the frequency of the VAA and thus are quite different from their weak coupling counterparts.

\begin{figure}[htbp]
\begin{center}
\includegraphics[width=8cm]{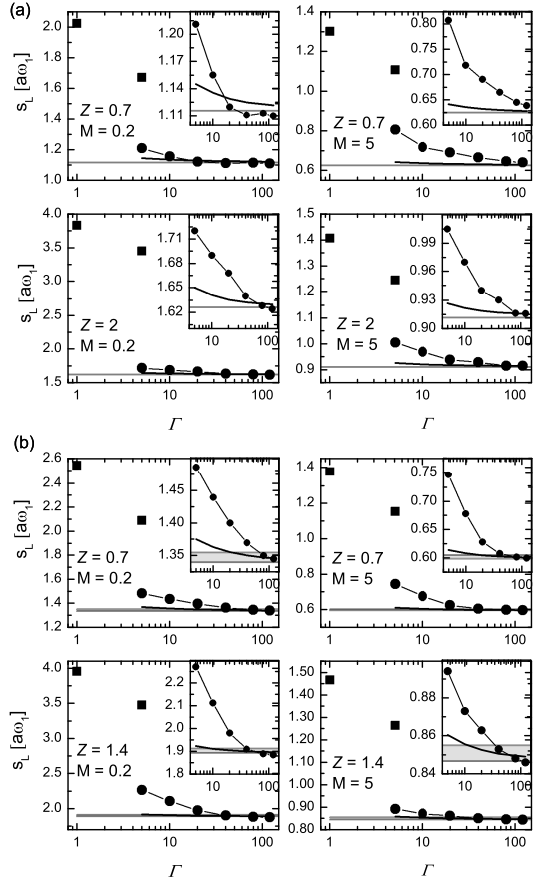}
\caption{Liquid state: Longitudinal sound speeds. (\ding{108}) MD, line: QLCA, gray shaded area/line: lattice value. For $\Gamma=1$ and $\Gamma=5$ the RPA (Vlasov equation based) values of the sound speeds are also indicated (\ding{110}). (a) $n_2=n_1/2$; (b) $n_2=n_1$.}
\label{fig:SoundS}
\end{center}
\end{figure}

\begin{figure}[htbp]
\begin{center}
\includegraphics[width=8cm]{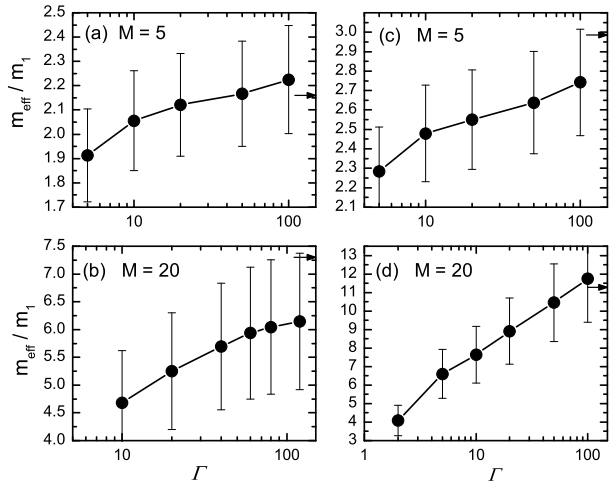}
\caption{Liquid state: Effective masses for $Z_1=Z_2$ vs. $\Gamma$. (a,b) $n_2=n_1/2$; (c,d) $n_2=n_1$. The arrows indicate the mass average $\langle m \rangle =(n_1m_1+n_2Z_2)/(n_1+n_2)$. The error bars represent 5\% (10\%) uncertainty in the measurement of the MD sound speed for $M=5$ (20).}
\label{fig:effmass}
\end{center}
\end{figure}

We have studied the $\Gamma$-dependence of the sound speeds and of the related effective masses, the latter being defined by subtracting the explicitly correlation dependent term from the sound speed coefficient
\begin{equation} \label{eq:efm}
\frac{m_{\rm eff}}{m_1} = \frac{\omega_1^2 a^2}{s^{L^2}}\frac{\langle Z \rangle^2}{\bar{\kappa}}\left[1+\frac{n_2}{n_1}\right](1-U)
\end{equation}
by MD simulations for the parameter set given previously. Results are shown in Figs. \ref{fig:SoundS} and \ref{fig:effmass} for $\Gamma$ values between $\Gamma=5$ and $\Gamma=120$. For $\Gamma$=1 and $\Gamma$=5 sound speed values calculated through the (Vlasov Equation based) RPA approach are also displayed. 

At the high $\Gamma$ end the QLCA predicted behavior is in excellent agreement with simulation results. As $\Gamma$ approaches the freezing boundary, the sound speeds smoothly join their values in the crystal lattice, which are also given, in anticipation of the results of the next Section. Some further comments on the relationship between the sound speeds in the two domains will be given there. In the liquid, as $\Gamma$ is lowered, the remarkable decrease of the effective mass and the concomitant increase of the sound speed can be observed. At the same time, the QLCA sound speed, in general, stays below the observed value because the QLCA ignores the modification of the effective mass as $\Gamma$ is reduced. It can be noted, that even at the relatively low $\Gamma=5$ value the strong coupling behavior seems to be still dominant and the sound speed is much below its calculated RPA value. The behavior of the sound speed below this $\Gamma$ value is not clear: it is a domain that would require substantial theoretical, simulation and experimental work to arrive at a reliable and coherent picture.

\subsection{Mode Dispersion}

Now we turn to the description of the full mode structure in the liquid state. By solving the characteristic equations for the matrices (\ref{eq:Clong}) and (\ref{eq:Ctran}) one obtains the full $\omega(k)$ dispersion for the 4 liquid modes. The results of this calculation are displayed for $n_2/n_1=1$ and $1/2$ density ratios and for the already chosen $Z=Z_2/Z_1=0.7$ and 1.4 (2.0), $M=m_2/m_1=0.2$ and $5.0$ parameter values. The $Z_2/Z_1$ values have been chosen in the vicinity of the stability boundary for the (staggered rectangular and honeycomb) binary lattices. 

Our theoretical analysis of the mode structure was accompanied by detailed Molecular Dynamics studies of the dynamical fluctuation spectra of the system, as described below.

In the Molecular Dynamics simulations we trace the trajectories of individual particles as obtained from the integration of the their equations of motion:
\begin{equation}
 m_i \frac{d {\bf v}_i}{dt}= - \sum_{j \neq i}^N \nabla \phi_{ij},
\label{eq:lmotion}
\end{equation}
where $\phi_{ij}$ is the interaction potential energy ($\propto Z_i Z_j$) between the particles $i$ and $j$, and $m_i$ are the masses of the particles. We use periodic boundary conditions, the edge lengths of the computational box ($L_x$ and $L_y$) and the total number of particles are chosen to accommodate a perfect lattice for the selected density ratios (and expected associated lattice structures). In the case of $n_2 / n_1 = 1$ we use $N_1=N_2=2040$ particles, while in the case of $n_2 / n_1 = 1/2$ we use $N_1=2720$ and $N_2=1360$ particles.

In the simulations of liquid-phase systems normally random initial particle configurations are set up. In all cases ample time is given to the systems to reach thermodynamic equilibrium before measurements on the systems start. During this equilibration phase, rescaling of the particle velocities is applied to reach the desired system temperature; this procedure is, however, stopped before data collection.

The central quantities to be calculated in the simulations are the fluctuation spectra of the densities and currents. Static pair distribution functions $g_{AB}(r)=1+h_{AB}(r)$ are also obtained and used as input for the QLCA calculations. Information about the (thermally excited) collective modes is obtained from the Fourier analysis of the correlation spectra of the density fluctuations of the different species ($A,B$= 1,2):
\begin{equation}\label{eq:rho}
\rho_A(k,t)= \sum_{j=1}^{N_\alpha} \exp \bigl[ i k x_j(t) \bigr],
\end{equation}
yielding the dynamical structure functions as \cite{HMP75}:
\begin{equation}\label{eq:sp1}
S_{AB}(k,\omega) = \frac{1}{2 \pi \sqrt{N_A N_B}} \lim_{\Delta t \rightarrow \infty}
\frac{1}{\Delta t}  \rho_A(k,\omega) \rho^\ast_B(k,\omega) ,
\end{equation}
where $\Delta t$ is the length of data recording period and $\rho(k,\omega) = {\cal{F}} \bigl[ \rho(k,t) \bigr]$ is the Fourier transform of (\ref{eq:rho}). The $(A,B)$ combinations label spectra related to component 1, $S_{11}$, to component 2, $S_{22}$, as well as to the cross term $S_{12}$.

Similarly, the spectra of the longitudinal and transverse current fluctuations, $L(k,\omega)$ and $T(k,\omega)$ are obtained from Fourier analysis of the microscopic quantities, respectively,
\begin{eqnarray} \label{eq:dyn}
\lambda_A(k,t)&=& \sum_{j=1}^{N_\alpha} v_{j x}(t) \exp \bigl[ i k x_j(t) \bigr], \nonumber \\
\tau_A(k,t)&=& \sum_{j=1}^{N_\alpha} v_{j y}(t) \exp \bigl[ i k x_j(t) \bigr],
\end{eqnarray}
where $x_j$ and $v_j$ are the position and velocity of the $j$-th particle. Here we assume that ${\bf k}$ is directed along the $x$ axis. These calculations allow the determination of the spectra for a series of wave numbers, which are multiples of $k_{min,x(y)} = 2 \pi / L_{x(y)}$, where $L_{x(y)}$ is the edge length of the simulation box in the $x$ (or $y$) direction.

The identification of the collective modes is based on the observation of the extrema of $L_{11}$ and $L_{22}$. When the peak positions do not completely coincide, (this may happen for various reasons, which will be discussed elsewhere), it is the position of the stronger peak that is accepted.

\begin{figure}[htbp]
\begin{center}
\includegraphics[width=8cm]{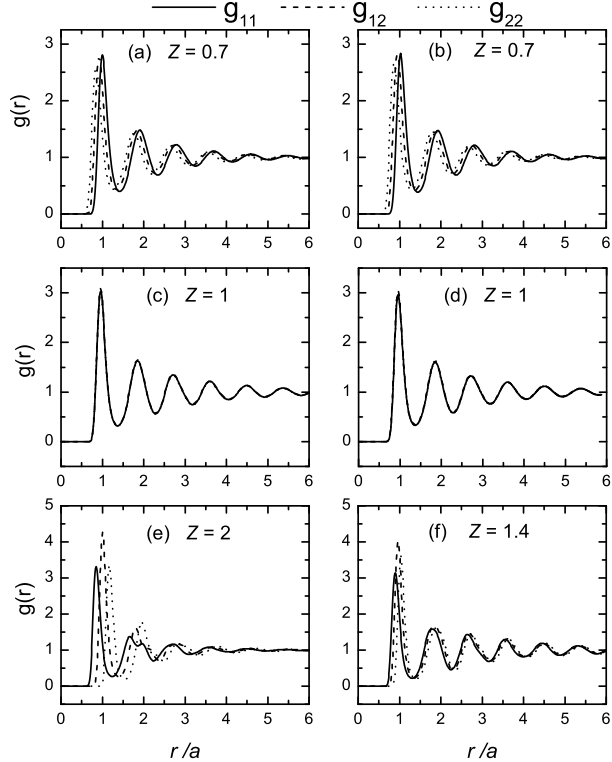}
\caption{Liquid state: Examples for the $g(r)$ pair distribution functions at $\Gamma=120$
(a,c,e) $n_2=n_1/2$; (b,d,f) $n_2=n_1$. Note that for $Z_1=Z_2$ we obtain $g_{11}=g_{22}=g_{12}$, irrespective of the density ratios.}
\label{fig:HCPCF} 
\end{center}
\end{figure}

Distribution functions $g_{AB}(r)=1+h_{AB}(r)$ that have been inputted in the QLCA calculations are given in FIG \ref{fig:HCPCF} for the previously chosen $n_2/n_1$ and $Z_2/Z_1$ values. We have also added the $Z_2/Z_1=1$ distribution functions, in order to show that in this case the three, $h_{11}$, $h_{12}$, and $h_{22}$, correlation functions are identical, independently of the density ratios (the mass ratios obviously do not affect the correlation functions).

\begin{figure}[htbp]
\begin{center}
\includegraphics[width=8cm]{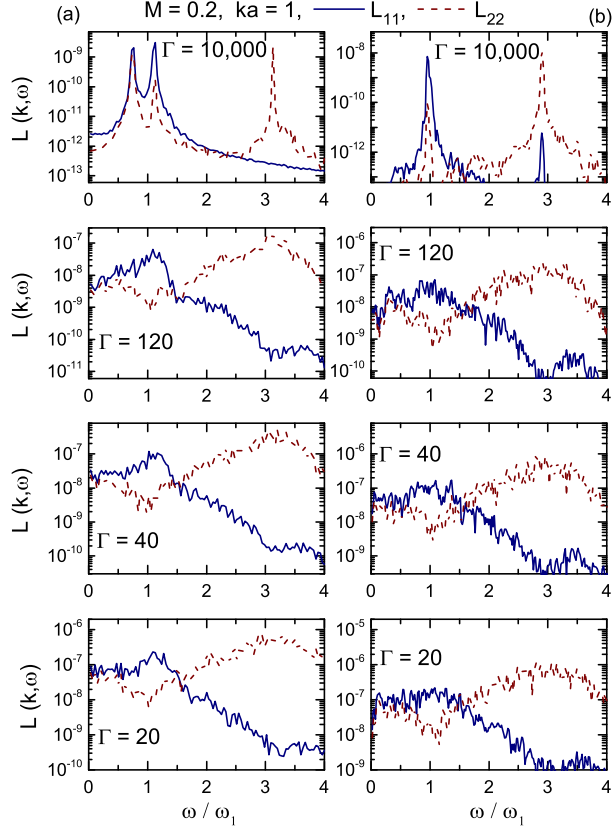}
\caption{Liquid state: Examples for the longitudinal current fluctuation spectra. (a) left column: $n_2=n_1/2$, $Z_2=2 Z_1$; (b) right column $n_2=n_1$, $Z_2=1.4 Z_1$.}
\label{fig:HCL11}
\end{center}
\end{figure}

Some illustrative current fluctuation spectra are given in Fig. \ref{fig:HCL11}.

\begin{figure}[htbp]
\begin{center}
\includegraphics[width=8cm]{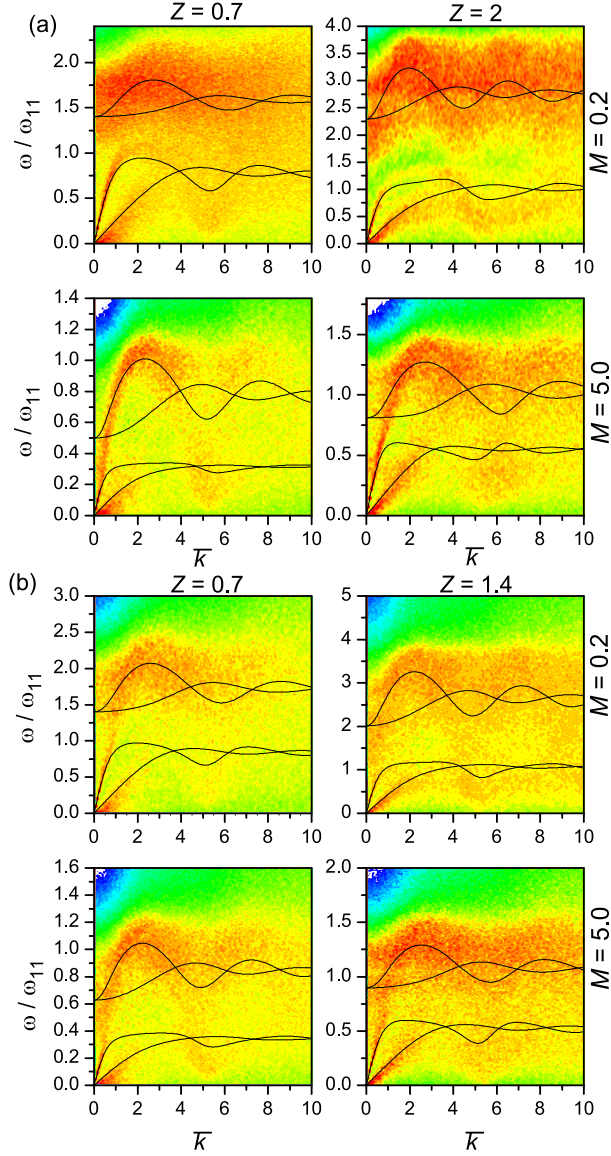}
\caption{Liquid state: Current fluctuation spectra from MD simulation (color map) compared with QLCA calculated dispersion (black lines) for $\Gamma=120$. (a) $n_2=n_1/2$; (b) $n_2=n_1$.}
\label{fig:MDQLCA}
\end{center}
\end{figure}

The MD simulated mode structures, together with the QLCA calculated dispersion curves are given in Fig. \ref{fig:MDQLCA}. Although the MD spectra are sometimes quite noisy as the collective modes have rather broad peaks in the spectra, the agreement between the simulated and calculated dispersions, in general, is good. A new feature shown by the simulation but not predicted by the QLCA formalism is the merging of a portion of the longitudinal acoustic and longitudinal optic modes at low $m_2/m_1$ values into a new acoustic mode.

\section{Binary Lattice}

Depending on the $Z$ and $n$ values of the two components, a variety of ordered and disordered phases should exist in a 2D binary crystal. In combination with the different melting temperature associated with the different phases, a rather complex phase diagram can emerge. The stability of the different structures can be analyzed through a thermodynamic approach \cite{LOWEN1,LOWEN2} (minimizing the free energy) or through a dynamical normal mode analysis. In this paper we restrict ourselves to the study of the $T=0$ ($\Gamma \rightarrow \infty$) lattice structures only, which is amenable to the latter approach.

The lattice calculation is based on the evaluation of the lattice sum for the dynamical matrix 
\begin{equation}
C_{AB}^{\mu\nu}({\bf k})=-e^2\frac{Z_AZ_B}{\sqrt{m_Am_B}} \left[ \sum_i\left\{\Psi^{\mu\nu}({\bf r}_{i,AB})\left(\exp(-i{\bf k}\cdot{\bf r}_{i,AB} - \delta_{AB}\right) - \delta_{AB} \sum_{C\ne A} \sum_j \frac{Z_C}{Z_A}\Psi^{\mu\nu}({\bf r}_{j,AC})\right\}\right]
\end{equation}
over all the particle pairs with designated $A,B$ and $A,C$ indices, which now run over all the bases in the primitive cell (the number of which may be equal to or greater than the number of species, i.e. 2). The evaluation was done for ca. $10^5$ particles.

In the following we will consider two different lattice structures with the previously studied density ratios $n_2/n_1=1$ and $n_2/n_1=1/2$. These two cases provide a reasonable guidance as to what lattice mode spectrum to expect in more general situations. In both cases we choose the equilibrium hexagonal lattice as the skeleton Bravais lattice. The descendent crystal structures should be stable in the vicinity of $Z_1=Z_2$. With $Z_1=1$, $Z_2$ is restricted to $Z_m<Z_2<Z_M$. The values of $Z_m$ and $Z_M$ have been determined by finding the onset of unstable normal modes \cite{stability} and are given for both cases in Table \ref{tab:1}. Then the resulting lattice structures are the following;

\begin{table}[htdp]
\caption{Stability regions.}
\begin{center}
\begin{tabular}{|c|c|c|}
\hline$n_2/n_1$ & $Z_m$ & $Z_M$ \\
\hline
1 & $0.646 \pm 0.001$ & $1.548 \pm 0.002$ \\
1/2 & $0.51 \pm 0.01$ & $2.88 \pm 0.01$ \\
\hline
\end{tabular}
\end{center}
\label{tab:1}
\end{table}

\begin{figure}[htbp]
\begin{center}
\includegraphics[width=8cm]{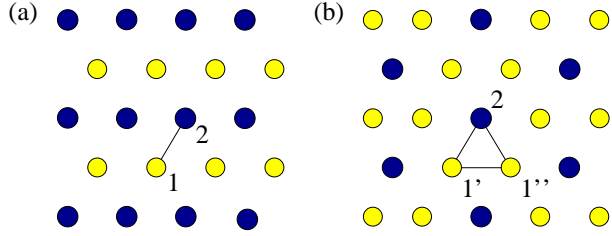}
\caption{Principal lattice structures: staggered rectangular (a), and honeycomb (b). Primitive cells are shown with dashed lines. In (b) the positions $1^\prime$ and $1^{\prime\prime}$ are distinguished.}
\label{fig:lattstruct}
\end{center}
\end{figure}

\begin{enumerate}
\item In the equal-density case with $n_1=n_2$ we obtain a staggered rectangular (SR) lattice with the aspect ratio $\sqrt{3}:1$.
\item In the half density case with $n_2=n_1/2$ we obtain a honeycomb (HC) lattice for species 1, while the particles of species 2 occupy the center sites of the honeycomb and form a hexagonal lattice; the lattice constants of the two lattices are in the ratio $\sqrt{3}:1$, see figure \ref{fig:lattstruct}.
\end{enumerate}

The SR structure is built up from 2 bases in the primitive cell, while the HC structure has 3 bases in the primitive cell. According to \cite{LOWEN1,LOWEN2} other possible structures may exist outside the stability domains of Table \ref{tab:1}, such as various rhombic structures, the asymmetric hexagon (also with 3 bases in the primitive cell) and various pentagonal structures (with 3-5 bases in the primitive cell).

The number of modes $r$ in general is $r= d \times b$, where $d$ is the dimensionality and $b$ is the number of bases in the primitive cell. In general, the polarizations of the modes can be characterized only in the combined $r$-dimensional species--configuration space. In specific situations, however, (i) the $r$-dimensional space factorizes into the $b$-dimensional species- and $d$-dimensional configuration sub-spaces; moreover, (ii) longitudinal and transverse polarizations (with respect to $k$) may become the eigen-polarizations in the latter. This occurs when ${\bf k}$ is along one of the principal axes of the crystal. Thus the $L_{+}$, etc. designations remain still meaningful and, by continuity, can be used for the labeling of the modes, with the proviso that since in general, more than one pair of optic modes may exist, a further index, say $\beta = I, II$ may be needed for the full labeling. The HC mode structure consists of 6 modes altogether, out of which 3 are ``longitudinal'' and 3 are ``transverse'' modes. Due to the rotational symmetry of the reciprocal lattice for $k \rightarrow 0$ the $L$ and $T$ optic modes are degenerate at $k=0$. In this limit, one can identify a pair of acoustic and two pairs of degenerate optic (gapped) modes. The SR mode structure consists of 4 (2 longitudinal and 2 transverse) modes. In the absence of rotational symmetry the $L$ and $T$ modes are not degenerate at $k=0$, and the $\omega_{-}^L$ and $\omega_{-}^T$ gaps are separated.

\subsection{Optic modes}

\begin{figure}[htbp]
\begin{center}
\includegraphics[width=8cm]{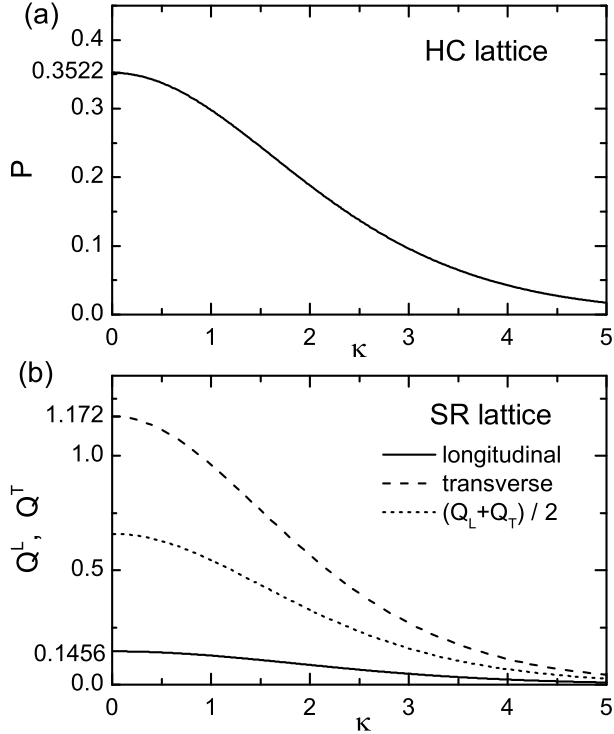}
\caption{The dependence of the lattice sums on the Yukawa screening parameter ($\kappa$). (a) HC lattice; (b) SR lattice.}
\label{fig:HCsumP}
\end{center}
\end{figure}

The simple geometric structure of the primitive cell allows one to obtain a transparent result for the $\omega(k\rightarrow 0)$ frequency gaps. The results are given below and portrayed in Fig. \ref{fig:HCsumP}. For the HC at $k=0$ the elements of the $C$-matrix are
\begin{eqnarray}\label{eq:OpC}
\frac{C_{1^\prime 1^\prime}^L(0)}{\omega_1^2}&=&\frac{1}{2\sqrt{2}}\left[\sum_j \Psi^L({\bf r}_{j,1^\prime 1^\prime})+\Psi^L({\bf r}_{j,1^\prime 2})\right]=\frac{C_{1^{\prime\prime} 1^{\prime\prime}}^L(0)}{\omega_1^2}  \\
\frac{C_{2 2}^L(0)}{\omega_1^2}&=&\frac{1}{2\sqrt{2}}\frac{Z}{m}\left[\sum_j \Psi^L({\bf r}_{j,21^\prime})+\Psi^L({\bf r}_{j,21^{\prime\prime}})\right] \nonumber \\
\frac{C_{1^\prime 1^{\prime\prime}}^L(0)}{\omega_1^2}&=&-\frac{1}{2\sqrt{2}}\left[\sum_j \Psi^L({\bf r}_{j,1^\prime 1^{\prime\prime}})\right]\nonumber \\
\frac{C_{1^\prime 2}^L(0)}{\omega_1^2}&=&-\frac{Z}{2\sqrt{2m}}\left[\sum_j \Psi^L({\bf r}_{j,1^\prime 2})\right] = \frac{C_{1^{\prime\prime} 2}^L(0)}{\omega_1^2} \nonumber
\end{eqnarray}

By symmetry, all the lattice sums are equal; the rotational symmetry ($L=T$) can be further exploited to obtain
\begin{equation}
P=\frac{1}{2\sqrt{2}}\sum_{j,21^\prime}\exp(-y)\frac{1}{\bar{r}^3}\left(\frac{1}{2}(1+y+y^2)\right)
\end{equation}
in terms of which the roots of the cubic equation are
\begin{eqnarray} \label{eq:PHCgap}
\omega^{L,T}_{-,I}& =& \sqrt{2\left(p^2+q^2\right)P} \\
\omega^{L,T}_{-,II}&=& \sqrt{2\left(1+p^2\right)P}, \nonumber 
\end{eqnarray}

Note that $\omega_{-,II}^{L,T}$ is an ``invariant mode'', where the gap frequency is independent of $m_2$; in this mode the light particles oscillate around the inert heavy particle.

For the SR a similar construction yields
\begin{equation}
Q^{L,T}=\frac{1}{2\sqrt{2}}\left[\sum_{j,12}\Psi^{L,T}({\bf r}_j)\right]
\end{equation}
in terms of which 
\begin{eqnarray} \label{eq:PSRgap} 
\bar{\omega}^L_{-}&=&\sqrt{\left(p^2+q^2\right)Q^L} \\ 
\bar{\omega}^T_{-}&=&\sqrt{\left(p^2+q^2\right)Q^T}. \nonumber 
\end{eqnarray} 

\begin{figure}[htbp]
\begin{center}
\includegraphics[width=8cm]{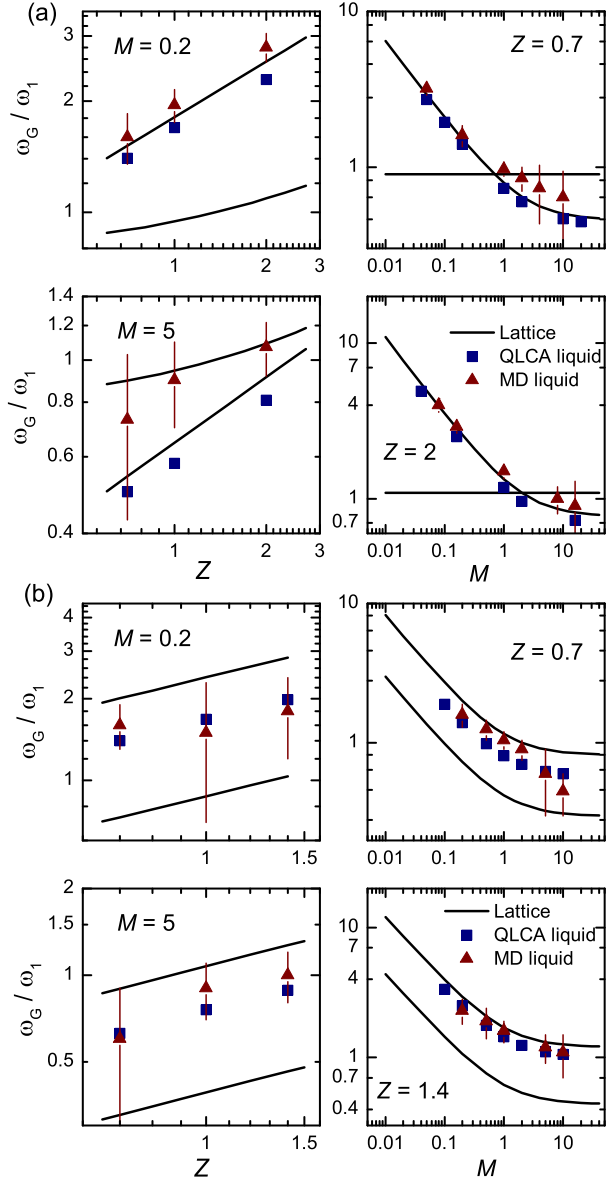}
\caption{Mass and charge ratio dependence of the gap frequencies. The QLCA and MD result are also shown. (a) HC lattice; note the portrayal of the ``invariant mode'' in the right ($Z={\rm const.}$) panels. (b) SR lattice.}
\label{fig:GapMZ}
\end{center}
\end{figure}

Here the $P$, $Q$-s are lattice sums, characteristic of the lattice structure (SR or HC); they depend on $\kappa$ only. They can be contrasted with $F_{12}$ factor appearing in the gap frequency expression in the liquid (\ref{eq:FFF}), that depends on $Z_2/Z_1$ as well (see Fig. \ref{fig:GapMZ}), but for $\Gamma \rightarrow \Gamma_{\rm freeze}$ its value in the $n_2=1/2 n_1$ and $n_2=n_1$ cases reasonably well approaches the corresponding $4P$ (for the HC) and $2(Q_L+Q_T)/2$ (for the SR) values respectively.

\begin{figure}[htbp]
\begin{center}
\includegraphics[width=8cm]{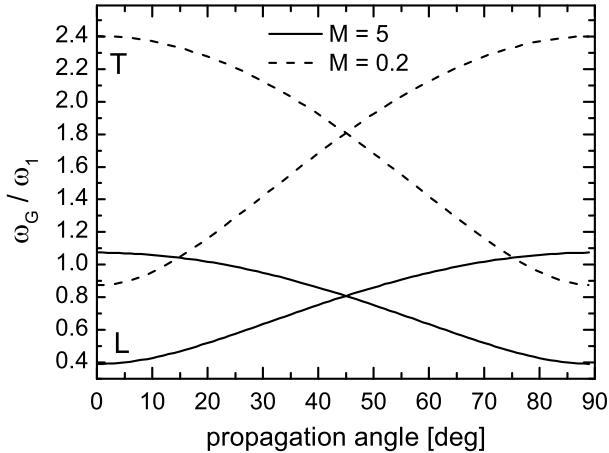}
\caption{SR lattice: polarizations of the gap frequencies versus propagation angle for $Z_2=Z_1$.}
\label{fig:SRgappol}
\end{center}
\end{figure}

While the gap frequencies are angle independent, the polarizations associated with them are not: Fig. \ref{fig:SRgappol} shows that $T$ and $L$ polarizations switch place as the propagation angle varies from 0 to 90 degrees.

Figure \ref{fig:GapMZ} shows the $Z$ and $m$ dependences of the respective gap frequencies in the SR and HC crystal lattices and the corresponding gap frequency in the liquid. Commenting on the HC case first, we note that the liquid has only one frequency gap and therefore there is no equivalent of the invariant mode in the liquid. Turning to the SR lattice, one observes the separation of the longitudinal $\omega_{-}^L$ and transverse $\omega_{-}^T$ gaps. The $\omega_{\rm GAP }$ frequency in the liquid largely follows the angular average of $\omega_1$ and $\omega_2$, but less closely than it does in the case of the YOCP \cite{IEEEH}.

Figure \ref{fig:HCsumP} shows the dependence of the $P$, $Q$ lattice sums on the screening parameter; the smooth extrapolation to the $\kappa=0$ value provides the input for the calculation of the noteworthy Coulomb gap frequencies via Eqs. \ref{eq:PHCgap} and \ref{eq:PSRgap}. In parenthesis we note that a little reflection shows that the $P(\kappa=0)$ value bears a close relationship to the $M=\sum r^{-3}$ dipole sum over a hexagonal lattice whose value is well-known \cite{GoldenPRB08}: $M/2=0.7985/b^3$ in terms of the Wigner-Seitz radius $b$. Then $P(\kappa=0)=2^{-13/4}(3^{3/2}-1)M/2$.

\subsection{Acoustic modes and sound speed}

In contrast to the optic modes, whose dispersion is highly structure dependent and is, in general, quite different from the corresponding mode dispersion in the liquid, the $k\rightarrow 0$ behavior of the acoustic phonons in the lattice is largely similar to that of their liquid counterpart. More precisely, the sound speeds, as calculated by the QLCA and verified by simulations, go over quite smoothly to the lattice sound speeds as $\Gamma$ crosses the freezing boundary. This is visible in Fig. \ref{fig:SoundS}. The only difference of some significance arises in the case of the SR lattice, due to the fact that its reciprocal lattice space is, in contrast to the HC  structure, anisotropic even in the $k\rightarrow 0$ limit. The most important observation, however, is that the notion of the VAA as a dominant feature for the low frequency excitations both in the liquid and in the solid state,  is of universal validity.

\subsection{Mode dispersion}

\begin{figure}[htbp]
\begin{center}
\includegraphics[width=8cm]{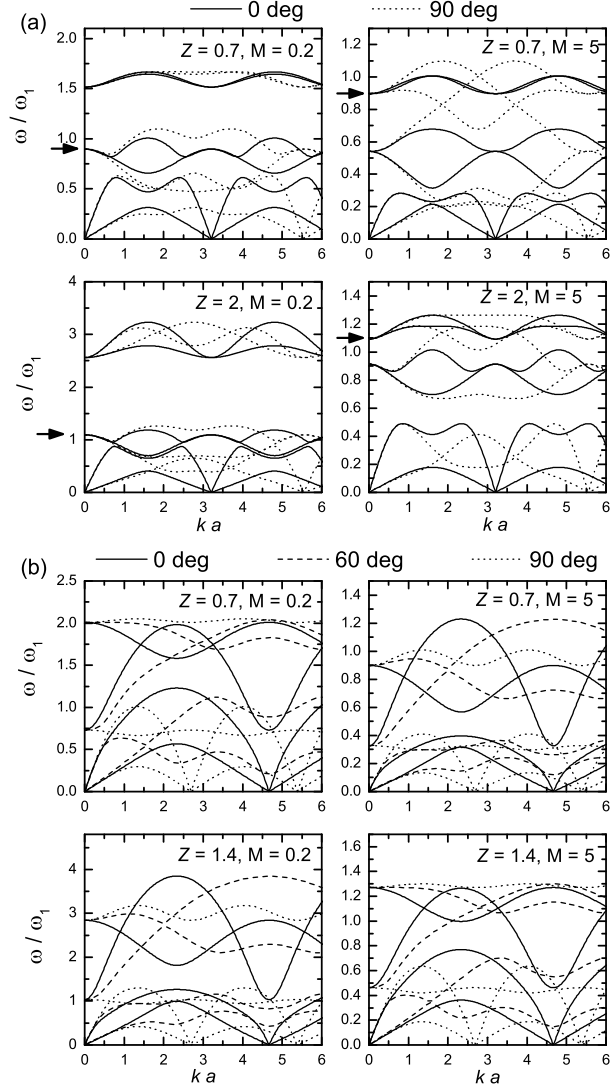}
\caption{Calculated mode dispersions for different propagation angles. (a) HC lattice, note that the invariant mode frequency at $k=0$ (pointed at by the arrows) remains invariant for any $M=m_2/m_1$ and any angle; (b) SR lattice.}
\label{fig:HCldisp}
\end{center}
\end{figure}

\begin{figure}[htbp]
\begin{center}
\includegraphics[width=8cm]{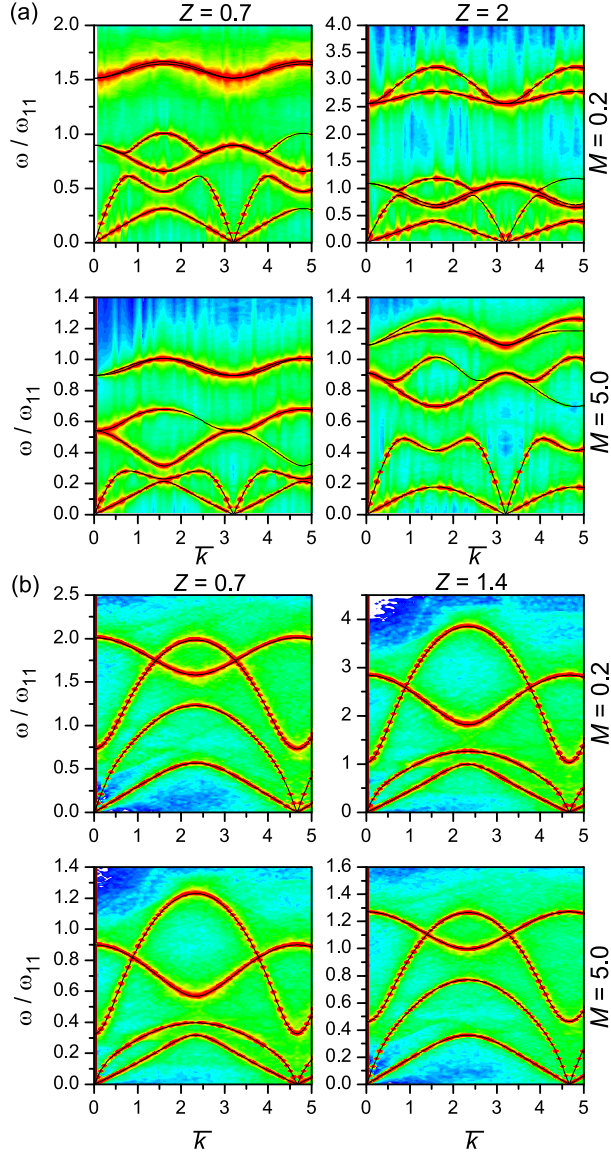}
\caption{Current fluctuation spectra from MD simulation at $\Gamma_1=10,000$ (color map) compared with dispersion from lattice calculations (black lines). (a) HC lattice; (b) SR lattice.}
\label{fig:MDlatt}
\end{center}
\end{figure}

The full calculated lattice phonon dispersion diagrams both for the HC and SR lattices are portrayed in Fig. \ref{fig:HCldisp}. In order to be able to compare the MD results with lattice summation data, simulations were carried out at very low temperatures, at $\Gamma_1 = 10^4$. In these runs the particles are initially arranged in a perfect lattice and their thermal motion does not disrupt the lattice in the course of the simulations. In Fig. \ref{fig:MDlatt} we display the MD simulation results for these finite temperature lattices: the MD simulations and the results of the lattice calculations  are in full agreement. 

\begin{figure}[htbp]
\begin{center}
\includegraphics[width=8cm]{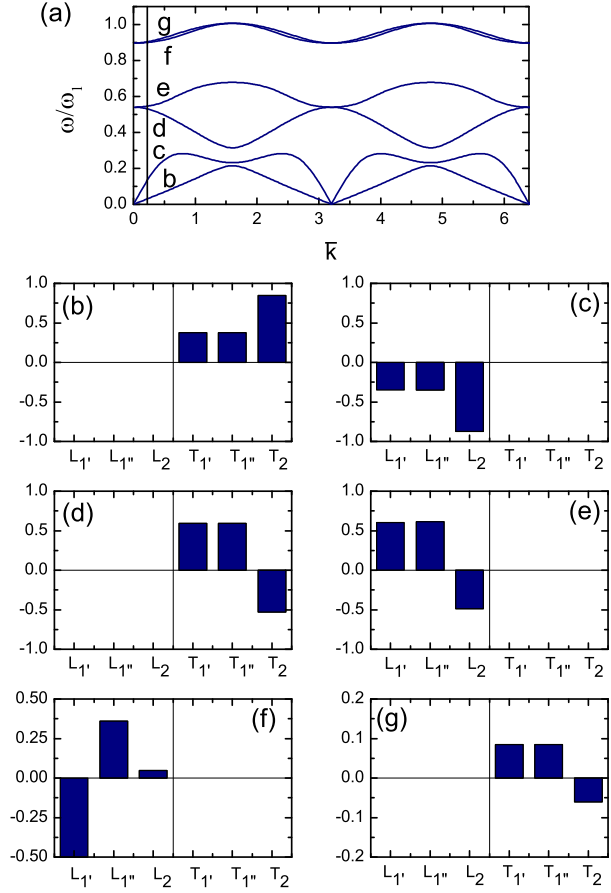}
\caption{HC lattice: Mode polarizations for 0 deg propagation. $Z = 0.7$, $M = 5$, $\alpha = 0^{\circ}$ ($k || x$), $ka = 0.2$ in the panels, particle "2" is the heavy one (see Fig. \ref{fig:lattstruct}).}
\label{fig:HCpol0}
\end{center}
\end{figure}

\begin{figure}[htbp]
\begin{center}
\includegraphics[width=8cm]{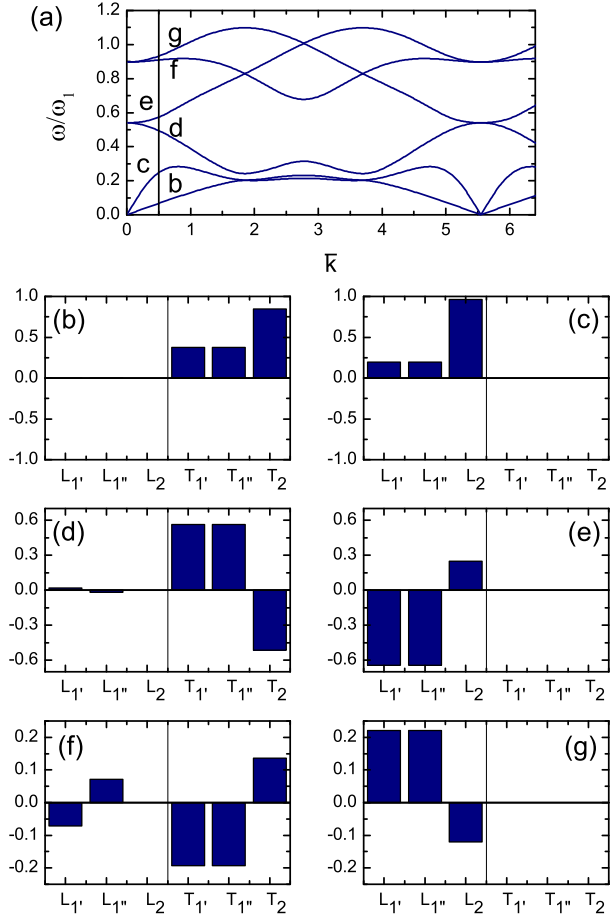}
\caption{HC lattice: Mode polarizations for 90 deg propagation. $Z = 0.7$, $M = 5$, $\alpha = 90^{\circ}$ ($k || y$), $ka = 0.5$ in the panels.}
\label{fig:HCpol90}
\end{center}
\end{figure}

\begin{figure}[htbp]
\begin{center}
\includegraphics[width=8cm]{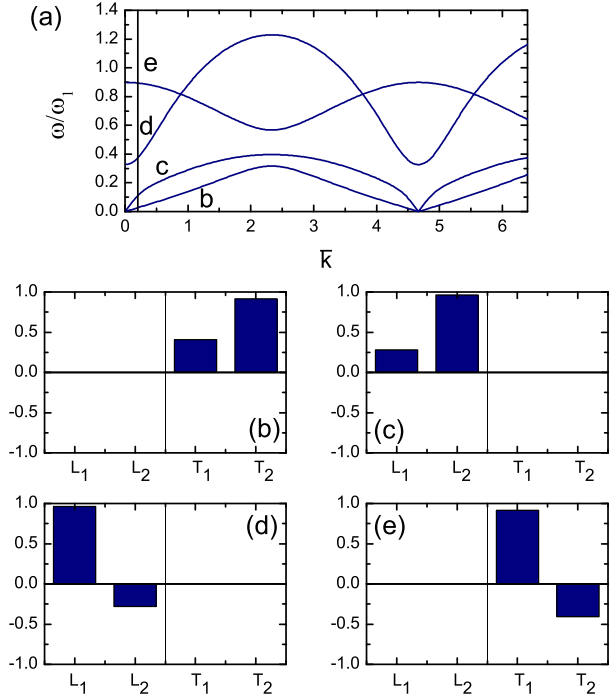}
\caption{SR lattice: Mode polarizations for 0 deg propagation. $Z = 0.7$, $M = 5$, $\alpha = 0^{\circ}$ ($k || x$), $ka = 0.2$ in the panels, particle ``2'' belongs to species ``2''. Note the $L/T$ and 1/2 polarization mixings.} 
\label{fig:SRpol0}
\end{center}
\end{figure}

\begin{figure}[htbp]
\begin{center}
\includegraphics[width=8cm]{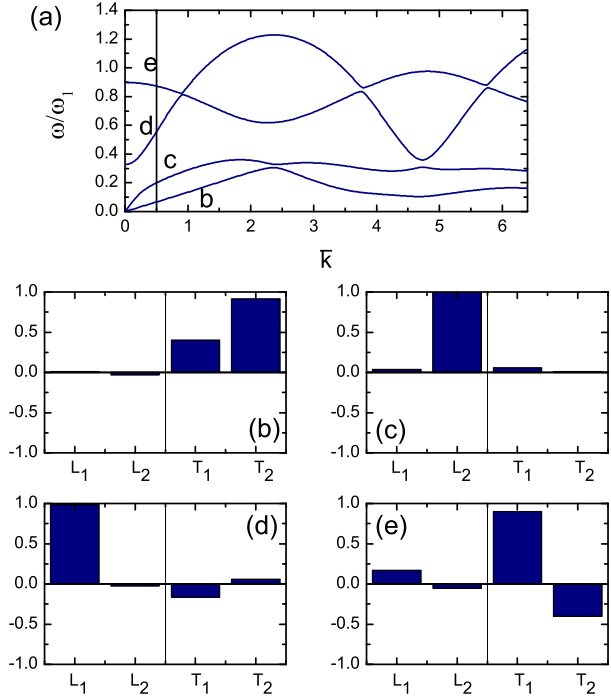}
\caption{SR lattice: Mode polarizations for 10 deg propagation. $Z = 0.7$, $M = 5$, $\alpha = 10^{\circ}$, $ka = 0.5$ in the panels, particle ``2'' belongs to species ``2''.}
\label{fig:SRpol10}
\end{center}
\end{figure}

The polarizations of the modes in the combined (in general not factorizable) cartesian and species space can be assessed from Figures \ref{fig:HCpol0}, \ref{fig:HCpol90} and \ref{fig:SRpol0}, \ref{fig:SRpol10}, where the components of the eigenvectors for the HC (SR) lattice modes along the 6 (4) eigendirections of the dynamical matrix for a given $k$ are shown. The lengths of the $L_A$ and $T_A$ labeled bars (components of the eigenvectors) are proportional to the longitudinal and transverse displacements of particles at position $A$. Samples are given for propagation angles along and off the principal axes. Note that in the latter case no overall polarization direction can be assigned to the displacement of the particles belonging to different species.

\begin{figure}[htbp]
\begin{center}
\includegraphics[width=8cm]{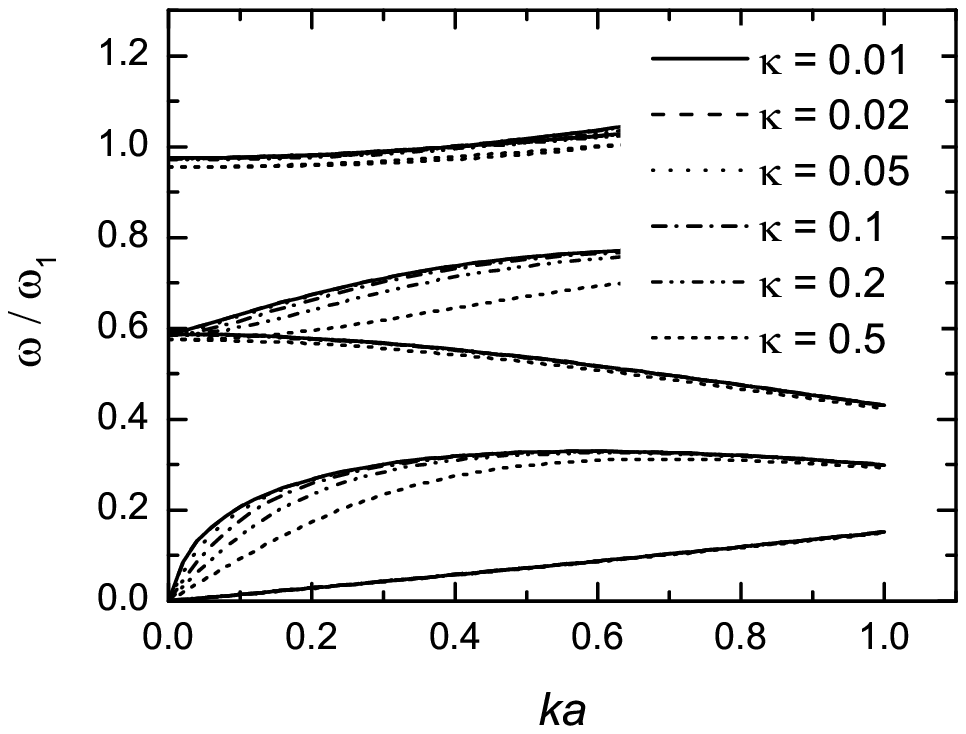}
\caption{The dependence of the HC lattice dispersions on the Yukawa screening parameter ($\kappa$) at $m_2=5m_1$ and $Z_2=0.7 Z_1$.}
\label{fig:HCkappa}    
\end{center}
\end{figure}

Finally we address the question of how the collective mode dispersion depends on $\kappa$, the screening parameter of the Yukawa potential and, in particular, how the transition to the $\kappa=0$ Coulomb limit occurs. Figure \ref{fig:HCkappa} shows that there is a smooth evolution of the mode dispersions towards the Coulomb limit and towards the changeover of the longitudinal acoustic mode into the characteristic quasiacoustic $\sqrt{k}$ Coulombic  behavior. It will be shown in another publication, that this behavior is in sharp contrast to what happens in the 3D case \cite{BIM}).

\section{Comparisons and Conclusions}

The results of earlier analyses \cite{DonkoJPC,PRL-Einstein,HartmannJPA06,IEEED,GoldenPRE10,KalmanEPL10} of the collective mode structure of the YOCP have established the close affinity of the mode structures in the strongly coupled liquid and in the crystalline lattice states. More precisely, what has been found is that the QLCA model, which essentially portrays the strongly coupled liquid as a superposition of randomly oriented microcrystals and determines the eigenmodes as those of the averaged crystal, provides an adequate description of wave propagation in the liquid. Whether such a simple picture would prevail in the binary liquid where the non-random distribution of the particles belonging to the two species is an issue as well, is not a priori obvious. The study presented in the previous Sections shows, however, that this is the case. In the following, we discuss the relationship between the phonon dispersion in the binary crystal, as calculated by lattice summation and corroborated by MD simulations, and collective excitations in the binary liquid, as provided by the QLCA description and the MD simulations. Judged by comparison with the results of the MD simulations, the QLCA results are quite reliable, with two exceptions, that we will discuss below. In comparing mode structures in the liquid and in the solid, the effect of the different density ratios has to be kept in mind: while in the former the difference between the $n_2=n_1$ and $n_2=n_1/2$ cases does not make a major difference, in the latter the two different crystal structures (SR and HC) substantially affect the mode structure.

Focusing first on the low-$k$ acoustic excitations, we see that there is an almost perfect agreement between the liquid QLCA and MD sound speeds, on the one hand, and the corresponding values in the liquid and in the two crystal structures studied, on the other hand. The only difference of note, as we have already pointed out, arises in the case of the SR lattice, due to its anisotropy, which results in a narrow band of sound speeds; in the liquid, as represented by the QLCA, it is replaced by an angular average. The most important result that emerges from all this is the fact that the low frequency excitations are governed by oscillation frequency $\omega$ of the Virtual Average Atom (see Eq. \ref{eq:226}) which is created by the average charges and masses of all the components. This effect, as it was discussed in some detail elsewhere \cite{PRL-2011}, has its most dramatic manifestation for the effective mass of the nominal plasma frequency of the binary (with respective masses $m_1$ and $m_2$), which in the weakly coupled case is formed, in general, through the ``parallel connection'' of the two masses ($1/m_{\rm eff}=1/m_1+1/m_2$), but which in the strongly coupled case becomes the ``series connection'' of the two masses ($m_{\rm eff}= m_1+m_2$). While the VAA has been a useful heuristic concept for liquid alloys \cite{VoraFMC08} and for disordered systems \cite{PoonPR66,ElliottRMP74,LangerJMP61,LikalterSCCS99}, and also in connection with self-diffusion \cite{HansenJoly}, here we have been able to give a rigorous demonstration through the QLCA of the emergence of this phenomenon. The MD simulation has shown (see Figs. \ref{fig:GapEins} -- \ref{fig:effmass}) that in the $\Gamma \rightarrow \Gamma_{\rm freeze}$ limit the VAA concept becomes ``exact'', in the sense that after the subtraction of the explicitly identifiable pair correlation $h_{12}$ dependent correlational contribution it determines the sound speed. With decreasing $\Gamma$ the $m_{\rm eff}$ decreases, seemingly marching towards weak coupling limit, but within the boundaries of our MD simulation which covers only the $\Gamma>5$ domain, the behavior is still essentially strongly coupled, in that the decrease of $m_{\rm eff}$ from its high $\Gamma$ value is quite slow. However, this decrease of $m_{\rm eff}$ in the moderately coupled domain is not reflected by the QLCA model: there $m_{\rm eff}$ preserves its high $\Gamma$ value (Eq. \ref{eq:efm}) for any $\Gamma$. This is the first inadequacy of the QLCA and it is the consequence of the fact that the appearance of the VAA structure is formally correlation independent. Correlational effects appear only indirectly, through the model from which it is derived and which adopts quasilocalization as its basis. That the quasilocalization can lead to such a qualitative effect is a novel feature of the approximation, which manifests itself only in binary systems. In contrast, in the single component system, the weakly coupled and strongly coupled states differ through their explicit correlation function dependence only.

A hallmark of the binary system is the emergence of -- one or more -- optic modes with a $k=0$ gap frequency. In the liquid state there is only one gap frequency, corresponding to the two -- longitudinal and transverse -- modes that become degenerate at $k=0$, due to the isotropy of the liquid. In the crystal lattice this degeneracy may or may not be lifted, depending on the local environment: it is in the SR crystal, but it survives in the HC crystal. In addition, in the crystal lattice the number of optic modes increases with the number of particles in the unit cell, which increases in oder to accommodate $n_2/n_1$ unequal 1 density ratios: hence an additional degenerate gap frequency in the HC crystal. This latter is the ``invariant mode'' whose gap frequency is independent of the mass of the lower density component, which remains inert in this mode. The mode does not have an equivalent in the liquid. The other, ``normal'' mode does re-appear in the liquid, with the gap frequency in the vicinity of the crystal equivalent (for the HC) or between the longitudinal and transverse gaps (in the SR). It should be emphasized though that the approximation of the liquid dispersion by angle averaging the lattice phonons is not equivalent to the QLCA. This difference was already demonstrated for the YOCP: here it is much more pronounced.

The gap frequencies are not related to the VAA. In the liquid they can be expressed in terms of the nominal Einstein frequencies $\bar{\Omega}_{AB}$ (Eq. \ref{eq:Qopt}) and thus they follow the ``parallel connection'' rule.

In the liquid state one can identify two upper ($\bar{\Omega}_{I}$) and lower ($\bar{\Omega}_{II}$) Einstein frequencies. (Eq. \ref{eq:EinF}) For high $k$ values the two ``acoustic'' (longitudinal and transverse) modes of the liquid merge into $\bar{\Omega}_{II}$, while the two optic modes merge into $\bar{\Omega}_{I}$. These latter cannot be directly identified in the crystal lattice, but they appear in the expressions for its gap frequencies, showing a good agreement with the QLCA calculated liquid $\bar{\Omega}_{I}$ and $\bar{\Omega}_{II}$ quantities.

According to the MD simulation result (Fig. \ref{fig:MDQLCA}) the slopes in the vicinity of $k=0$ of the longitudinal acoustic and longitudinal optic modes match and the two modes fuse into a single acoustic mode. There is no indication of this phenomenon within the QLCA formalism.

As to the dependence on the screening constant $\kappa$, we see (Fig. \ref{fig:HCkappa}) that the qualitative features of the dispersion remain unaffected over a wide range of $\kappa$ values, down to and including the $\kappa=0$ Coulomb limit.

A number of problems relating to the collective dynamics of the system have been identified, but have not been studied in this paper: the damping of the modes, the detailed structures and the link between the various fluctuation spectra, the nature of the underlying order in the liquid phase, lattice stability and structures, etc. These problems will have to be investigated in future works.

\begin{acknowledgments}
This work was supported by NSF Grants PHY-0715227, PHY-1105005, PHY-0812956 and the Hungarian Fund for Scientific Research (OTKA) through grants K77653, IN85261, K105476, NN103150.
\end{acknowledgments}


%

\end{document}